\documentstyle[prl,aps,multicol]{revtex}
\begin{document}
\title{\bf Superconductivity and Antiferromagnetism: Hybridization 
Impurities in a Two-Band Spin-Gapped Electron System} 
\vspace{1.0em}
\author{A.~ A.~Zvyagin$^{(a,b)}$ and H.~Johannesson$^{(a)}$}
\address{$^{(a)}$Institute of Theoretical Physics, Chalmers 
University of Technology and G\"oteborg University, SE 412 96 
G\"oteborg, 
Sweden}
\address{$^{(b)}$B.~I.~Verkin Institute for Low Temperature Physics 
and Engineering \\ of the National Ukrainian Academy of Sciences, 
310164 
Kharkov, Ukraine}
\date{\today}
\maketitle
\begin{abstract}
We present the exact solution of a one-dimensional model of a spin-
gapped correlated electron system with hybridization impurities 
exhibiting 
both magnetic and mixed-valence properties. The host supports 
superconducting fluctuations, with a spin gap. The localized 
electrons create a band of antiferromagnetic spin excitations {\em 
inside the gap} for concentrations $x$ of the impurities below some 
critical 
value $x_c$. When $x\!=\!x_c$ the spin gap closes and a 
ferrimagnetic 
phase appears. This is the first example of an exactly solvable model 
with 
coexisting superconducting and antiferromagnetic fluctuations which 
in addition supports a quantum phase transition to a (compensated) 
ferrimagnetic phase. We discuss the possible relevance of our results 
for 
experimental systems, in particular the U-based heavy-fermion 
materials. 
\end{abstract}
\centerline{PACS numbers 74.62.Dh, 75.30.Mb, 74.25.Ha, 75.20.Hr}

\begin{multicols}{2}
\narrowtext

There has recently been a renewed interest in the role of impurities 
in spin-gapped electron systems. Examples include magnetic or mixed-
valent impurities in ordinary BCS-like superconductors \cite{BT}, 
heavy 
fermion alloys \cite{hf}, and underdoped high-$T_c$ cuprates 
\cite{Lee}, as 
well as disorder driven quantum phase transitions in superconducting 
thin 
films \cite{GM}. While many effects from impurities in these systems 
are 
well understood, such as the pair breaking in an ordinary 
superconductor 
due to the violation of time-reversal symmetry by magnetic impurities 
\cite{sup}, others remain to be clarified --- the coexistence of 
superconductivity and antiferromagnetism in certain U-based heavy 
fermion 
compounds \cite{hfs,rechf} being a prime example.  

In this paper we study an exactly solvable model of a {\em finite 
concentration} of hybridization impurities in a one-dimensional (1D) 
multiband correlated electron host with a spin gap. Our choice of a 
multiband model is in part inspired by recent experiments on the 
heavy 
fermion compound U$_{1-x}$Th$_x$Be$_{13}$, showing a scattering of 
electrons off impurities in {\em two} channels \cite{Al1}. Equally 
important, multiband electron models are intrinsically interesting, 
in 
particular when allowing for interactions with defects, when they 
generically exhibit non-Fermi liquid behavior \cite{SS}. Here we 
address 
the question how this behavior gets influenced by electron 
correlations 
in a {\em spin-gapped} host.   
 
The exact solution of our model --- to be defined below --- reveals 
several remarkable properties. Most interestingly, for a finite 
concentration of impurities, too weak to destroy the spin gap in the 
host, there is an impurity-induced band of antiferromagnetic spin 
excitations living {\em inside the gap}, suggestive of a coexistence 
of 
superconducting and antiferromagnetic fluctuations. As the impurity 
concentration increases, the gap closes, signaling a quantum phase 
transition to a fully compensated ferrimagnetic phase. To the best of 
our knowledge, ours is the first exactly solvable model where 
antiferromagnetic fluctuations of localized electrons {\em and} 
superconducting fluctuations of itinerant electrons are taken into 
account simultaneously, allowing for a fully non-perturbative 
analysis of the relevant physics. 

As Hamiltonian for the electron host we take \cite{Sch}: 
\begin{eqnarray}
&&{\cal H}_{host} = \int dx \biggl(- 
\psi^{\dagger}_{m\sigma}(x)\partial_x^2 
\psi_{m\sigma}(x) + c \int dx' \nonumber \\
&&\,\times \delta (x-x')\psi^{\dagger}_{m\sigma}(x)
\psi^{\dagger}_{m'\sigma'}(x')\psi_{m'\sigma}(x')\psi_{m\sigma'}(x)
\biggr)
\ , \ \label{host}
\end{eqnarray}
where the electron fields $\psi_{m\sigma}$ carry band indices $m\!=\!1,2$ and 
spin indices $\sigma \!=\!\pm 1/2$, with repeated indices in 
(\ref{host}) summed over. The screened electron-electron interaction 
is 
here simulated by a local {\em exchange} potential with strength $c > 
0$, 
with electrons in the same band (and opposite spin) experiencing an 
attraction while electrons in different bands (with the same spin) 
repel 
each other. This interaction, which is integrable \cite{Sch}, 
produces 
Cooper-like singlet pairing bound states at any temperature, but 
since 
the model is one-dimensional there is no global phase coherence 
between 
pairs, and hence no long-range order. However, the pairing implies a 
critical field $H_c$ at zero temperature below which there is no 
magnetic response, reminiscent of the Meissner effect (although there
is of course no {\em manifest} Meissner effect in the system). 

Next, we introduce a Hamiltonian term describing a (band-neutral) 
hybridization of itinerant electrons with impurities, with $N_i$ 
counting their number \cite{Sch1}:
\begin{eqnarray}
&&{\cal H}_{imp-el} = \sum_j^{N_i}\sum_{m,\sigma,M}[{\theta\over 4}
|SMm\rangle \langle SMm| + 
\nonumber \\
&&\sum_{M'}A_{M,M'}(\psi^{\dagger}_{m\sigma}(x_j)|SMm\rangle \langle 
S'M'm| + H.c.)] \ .
\label{imp}
\end{eqnarray}
Here $|SMm\rangle$ denotes an impurity state with spin $S$ and 
projection $M$ ($S'\!=\!S\!-\!1/2 \ge 0$) satisfying the constraint 
$\sum_{M'}|S'M'm\rangle \langle S'M'm| + \sum_{M}|SMm\rangle 
\langle SMm| = 1$, $\theta$ measures the off-resonance shift of the 
impurity from the Fermi level, $A_{M,M'}^2 = -2Sc(M\sigma|M')^2$ with 
the Clebsch-Gordan coefficients selecting states with 
$M'\!=\!M\!+\!\sigma$ as $(M\sigma|M') \equiv (S{1\over2};M\sigma
|S(S-{1\over2});{1\over2}M')$. Each impurity can temporarily absorb the 
spin of an itinerant electron to form an effective spin 
$(S-{1\over2})$, i.e. the impurity appears in {\em two spin configurations} 
\cite{Sch1}. Note that the spin sector of ${\cal H}_{imp-el}$ 
corresponds to a local antiferromagnetic exchange 
$2c{\mbox{\boldmath $S$}}_j\cdot{\mbox{\boldmath $\tau$}}(x)$ 
between the impurity spin ${\mbox{\boldmath $S$}}_j$ and the electron 
spin density ${\mbox{\boldmath $\tau$}}(x)$. The two-particle 
scattering matrices of the model in (\ref{host}) and (\ref{imp}) 
satisfy the Yang-Baxter relations, thus providing for complete 
integrability \cite{KBI}. The model remains integrable for any 
distributions of $S_j$ and $\theta_j$ but for simplicity we here 
focus 
on the case where all impurities carry the same spin, with an equal 
number of positive and negative shifts $\pm \theta$. 

The energies and eigenstates of the model are parameterized by three 
sets of {\em rapidities}: charge rapidities $\{k_j\}_{j=1}^N$ (with 
$N$ the number of itinerant electrons), spin rapidities 
$\{\lambda_{\alpha}\}_{\alpha=1}^M$ (with $M$ the number of ``down 
spins''), and band rapidities $\{\xi_{\beta}\}_{\beta=1}^n$ (with 
$n$ the number of electrons in the $m=1$ band). (Note that a 
crystalline field will lift the degeneracy in Eq.~(\ref{host}), with 
the bands becoming unequally populated.) Each eigenstate corresponds 
to a solution of the Bethe {\em Ansatz} equations, here obtained on a 
periodic interval of length $L$, 
\begin{eqnarray}
&&e^{ik_jL} = \prod_{f=1}^M e_1(k_j - \lambda_f)\prod_{q=1}^n 
e_1^{-1}(k_j - \xi_q) \ , \nonumber \\
&&\prod_{\pm}e_{2S}^{N_i/2}(\lambda_{\alpha} \pm \theta)\prod_{j=1}^N
e_1(\lambda_{\alpha} - k_j) = - \prod_{f=1}^M e_2(\lambda_{\alpha} - 
\lambda_f) \ , \nonumber \\
&&\prod_{j=1}^N e_1(\xi_{\beta}- k_j) = - \prod_{\gamma=1}^n 
e_2(\xi_{\beta} - \xi_{\gamma}) \ , \
\label{BAE}
\end{eqnarray} 
where $e_n(y) = (2y-inc)/(2y+inc)$, $j = 1, \ldots, N$, $\alpha = 1, 
\ldots M$, and $\beta = 1, \ldots, n$, and with the energy given by 
$E=\sum_{j=1}^Nk_j^2$. To model a RKKY interaction in a real system 
we add an antiferromagnetic coupling between neighboring impurities 
\cite{fin}. While Eqs.~(\ref{BAE}) are insensitive to this addition, 
it does cause a shift of the energy: $E \to E - 
2\pi x\sum_{\alpha}^{N_i}a_{2S}(\lambda_{\alpha}-\theta)$ where 
$x=N_i/L$ is the impurity concentration, and $a_n(y)$ is the 
Fourier transform of $\exp (-n|pc|/2)$. The structure of the 
impurity-impurity interaction is similar to that of Eqs.~(\ref{host}),
(\ref{imp}), but with impurity state operators replacing electron 
fields. In the case when {\em all} electrons are localized it 
collapses to the well-known Takhtajan-Babujian spin-exchange 
Hamiltonian \cite{TB}. The fact that energies and eigenstates are 
blind to the spatial distribution of impurities is an artifact of 
integrability. One should note, however, that real systems, 
including certain heavy fermion alloys and dopped superconductors 
\cite{hf,hf1}, often exhibit a large quasi-degeneracy of the states 
as function of the distribution of impurities for stoichiometric 
compounds. Seen from this perspective, the {\em exact} degeneracy in 
our model can be turned to an advantage.  

Within the framework of the standard ``string hypothesis'' 
\cite{KBI}, the rapidities satisfying (\ref{BAE}) fall into four 
classes in the thermodynamic limit (with continuous distributions): 
(i) real charge rapidities of unbound itinerant electrons; (ii) 
pairs of complex conjugated charge rapidities representing 
Cooper-like singlet spin pairs; (iii) spin bound states 
($\lambda$-strings); and (iv) inter-band bound states ($\xi$-
strings). Since in 1D the most interesting behavior appears in 
the groundstate (corresponding to the low-temperature phases of 
higher-dimensional analogs) we focus on this in the following. 

The groundstate is obtained by filling up the Dirac seas of the low-
lying excitations, i.e. by populating all possible states with 
negative energies. An analysis shows that when $x=0$ only unbound 
electrons, Cooper-like spin-singlet pairs, and inter-band strings of 
lengths 1 and 2 have negative energies, and thus make up the Dirac 
seas defining the groundstate. The introduction of a finite 
concentration $x \neq 0$ of hybridization impurities drastically 
affect the groundstate (and in fact all states that are effectively 
present at a temperature $T \ll H$ in the case that an external 
magnetic field $H$ is applied): The spin levels of the localized 
electrons now form a low-lying {\em band of spin excitations 
inside the spin gap of the host}, given by spin strings of length 
$2S$. To see how this comes about, we study the integral equations 
for the dressed energies of the excitations (where the "dressing" 
is due to interactions \cite{KBI}), which, in the case of 
degenerate bands (zero band splitting) can be done analytically. 
We first integrate the equations for the dressed energies of the 
inter-band excitations, as these do not feel the presence of the 
impurities. The resulting set of equations for the dressed energies of 
unbound electrons ($\varepsilon$), Cooper-like pairs ($\psi$), 
and spin $2S$-strings ($\phi_{2S}$) has the form:
\begin{eqnarray}
&&(1-G_1)*\varepsilon = k^2 - \mu - {H\over2} - G_0*\psi + 
a_{2S}*\phi_{2S} \ , \nonumber \\
&&\psi =  2(\lambda^2 -{c^2\over4} - \mu) - G_0*\varepsilon  - 
x \pi \sum_{\pm}a_{2S}(\lambda \pm \theta) \ , \nonumber \\
&&W*\phi_{2S} = 2SH - x V - a_{2S}*\varepsilon  \ , \
\label{dren}
\end{eqnarray}
where $\mu$ is the chemical potential, $*$ denotes the convolution 
over the appropriate Dirac seas, $G_n(y)$ and $W(y)$ are the 
Fourier transforms of $\exp (-n|pc|/2)/2\cosh (pc/2)$ and 
$\coth (pc/2)[1\!-\!\exp (-2S|pc|)]$, respectively, and $V(p) = 
\cos (p\theta) \coth (pc/2)[G_0(p)\!-\!G_{4S}(p)]$. As revealed by 
(\ref{dren}), for $H\!=\!0$ and for sufficiently small 
concentrations of the impurities, the Dirac sea of spin-$2S$ strings 
is filled completely. The value of the spin gap (the gap for unbound 
electron excitations) is the smallest energy required to depair the 
Cooper-like spin-singlet state. The gap renormalizes in the presence 
of a small finite concentration of hybridization impurities, and we 
find that 
\begin{eqnarray}
&&\Delta (x) = \Delta (0)-{x \over 2} \sum_{\pm}\bigl( 
\int_{|\lambda| \ge Q}{d\lambda~s~{\rm sech} (\pi \lambda /c) 
\over (\lambda \pm \theta)^2 + (cS)^2} 
\nonumber \\
&&+ \frac{cS}
{(Q \pm \theta)^2 + (cS)^2}[(2/\pi ) \tan^{-1} \sinh (\pi Q/c) - 
1]\bigr) \ , \ 
\label{gap}
\end{eqnarray}
where $Q$ is the Fermi level for paired electrons. The spin gap 
decreases due to localized electrons (with their number given by 
$(N_i/ \pi)\sum_{\pm}\tan^{-1}[(Q \pm \theta)/ cS]$) for large 
$\theta$ and {\em closes} at a critical concentration $x_c$ (i.e. 
$\Delta(x_c) =0$). As seen from (\ref{gap}), the smallest value of 
$x_c$ is obtained when $S = 1/2$, with $x_c$ increasing with 
increasing $S$. The gap persists for larger impurity concentrations 
when the number of paired electrons ($\propto Q$) is large. 
However, and this is important, there are no additional unbound 
electron excitations 
appearing when the gap is open (i.e. for $x < x_c$). Hence, 
the presence of the hybridization impurities {\em does not lead 
to a pair-breaking} for these concentrations, in contrast to the 
suppression of superconductivity in ordinary BCS-like 
superconductors \cite{sup}. The presence of the spin 2S-strings 
in the gapped phase indeed suggests a coexistence of 
antiferromagnetic and superconducting fluctuations. It is 
important to note that the charge and magnetic subsystems are 
effectively {\em disconnected} for $x < x_c$, as revealed by 
Eqs.~(\ref{dren}): The low-energy (conformal) limit corresponds 
to a direct sum of {\em free bosons} (hard-core pairs) with 
scaling dimension $\eta=1$, and antiferromagnetic {\em spin 
strings} with $\eta = r(r+2)/4(S+1)$, $r=1,2,\dots$ (spinons 
with the minimal $\eta=1/2$ when $S=1/2$) \cite{AM}. 

The gap, and hence the energies of unbound electrons, may become 
negative when $x > x_c$. Thus, a Fermi sea for unbound electron 
excitations (with $k_F \propto \sqrt{x-x_c}$) opens up {\em in 
the absence of a magnetic field}, signaling a {\em quantum phase 
transition}. For $H\!=\!0$ there are no holes in the distribution 
of spin $2S$-strings, and hence the total magnetization of the 
system remains zero: The appearance of itinerant electrons 
correlates with the number of localized electrons, with both 
particle numbers scaling with $\sqrt{x-x_c}$, thus producing a 
{\em compensated ferrimagnetic phase} of the total system. (Note 
that there are two kinds of magnetic excitations in the system: 
spin $2S$-strings and unbound electrons which carry spin $1/2$.) 
This compensating effect is due to the RKKY type interactions 
between the impurities (which were absent in the attractive 
Hubbard model with hybridization impurities recently studied in 
\cite{fin}, where instead a ferromagnetic phase appears). 
However, Cooper-like pairs are still present for $x > x_c$, 
reminiscent of {\em gapless} superconductivity, where the gap 
closes before superconductivity is destroyed \cite{sup}. We here 
note that this is different from the supersymmetric $t\!-\!J$ 
model with hybridization impurities \cite{fin}, since for that 
model all low-lying excitations are gapless also with no 
impurities present. 

Let us briefly discuss the model in the presence of a magnetic field 
$H$. For $H$ smaller than some critical field $H_c(x)$, the spin gap 
persists for $x < x_c$, and there are no unbound electrons in the 
system. The critical field $H_c(x)$ decreases with increasing 
concentration $x$ and vanishes when $x$ approaches $x_c$. Thus, as 
expected, the spin gap decreases with the growth of a magnetic field 
and the concentration of impurities. The critical line separating 
the gapless and gapfull (ferrimagnetic) phases manifests itself in 
the van Hove singularity of the opening of the band of unbound 
electron excitations. The critical behavior at the {\em quantum 
phase transition} point $x\!=\!x_c, H\!=\!0$ also shows up in the 
scaling dimensions: Since an additional band of unbound electron 
excitations now appears, the dressed charge matrix \cite{KBI} gets 
enlarged from $2\times 2$ to $3\times 3$ with the off-diagonal 
components proportional to $\sqrt{x-x_c}$. The emergence of 
$2S$-spin strings when $S > 1/2$, together with the fact that their 
bandwidth is relatively narrow, $\propto x$, see Eqs.~(\ref{dren}), 
for this case imply the appearance of an additional critical 
magnetic field $H_{c1}$ at which the band of spin strings becomes 
empty (provided that the concentration of impurities is finite, 
since for $x\!=\!0$ $H_{c1}\!=\!H_c$). In addition, there is the 
possibility of an additional quantum phase transition for $H=0$ at 
$x_{c1} > x_c$, where the band of spin strings may become empty, 
with the impurities spin-polarized, analogous to the case of the 
attractive Hubbard chain with noninteracting hybridization 
impurities \cite{fin}. The additional band (of spin strings) 
{\em increases the effective mass} of the electrons, producing a 
larger coefficient for the low-temperature specific heat 
($\propto T$). In contrast, at the critical lines, the van Hove 
singularities of empty bands (of unbound electrons or spin 
strings) produce a $\sqrt{T}$ behavior of the specific heat.  

To summarize, in this paper we have studied an exactly solvable model 
of spin$-S$ hybridization impurities embedded into a two-band 
correlated electron host with a spin gap. A finite fraction of the 
electrons localize, producing a mixed valence of the impurities. 
These magnetic impurities interact, giving rise to a band of 
low-lying antiferromagnetic excitations appearing inside the spin gap 
of unbound conduction electrons. The simultaneous presence of Cooper 
pairs and spin density waves {\em in the spin gap} suggests that 
superconducting and antiferromagnetic fluctuations in fact coexist in 
this system. This is an important result, more so since it is 
obtained by an exact method taking {\em all} possible fluctuations 
into account, with no {\em a priori} assumption of a (local) symmetry 
breaking. We predict that the spin gap closes for concentrations of 
impurities above some critical value, at which unbound itinerant (and 
localized) electrons appear. However, the total magnetization of the 
system remains zero for $H\!=\!0$, implying a quantum phase 
transition to a compensated ferrimagnetic phase. A nonzero magnetic 
field less than the critical field $H_c$ does not destroy the 
coexistence of superconducting and antiferromagnetic fluctuations 
(provided that also the impurity concentration is smaller than the 
critical concentration $x_c$). 

Finally, let us briefly discuss the possible relevance of our results 
for real multiband systems with hybridization impurities. It is clear 
that the one-dimensionality of the model --- required for its exact 
solvability --- introduces features not seen in higher dimensions. In 
particular, spin-charge separation, implying no response of the 
paired 
system to an external field coupled to the current operator, is 
believed to be an intrinsically one-dimensional phenomenon. On the 
other hand, our model shares a number of important characteristics 
with real materials such as the U-based heavy fermion compounds, the 
alloy U$_{1-x}$Th$_x$Be$_{13}$ \cite{hfs,rechf,Al1} being a case in 
point. Common properties include (i) an enhancement of the 
effective mass \cite{Fel}; (ii) non-Fermi-liquid behavior in the 
normal (non-gapped) phase \cite{Ram}; (iii) power-law behavior of 
the low-temperature specific heat in the presence of a spin gap 
\cite{Fel}; (iv) scattering of the electrons off two configurations 
of 
the impurity ion \cite{Al1,SAC}; (v) the presence of low-lying 
magnetic excitations in the superconducting as well as in the normal 
state \cite{Ram}; (vi) coexistence of superconducting and 
antiferromagnetic fluctuations for some concentrations of magnetic 
impurities (U ions) \cite{hfs,rechf}; (vii) a (ferrimagnetic) phase 
with a weak ferromagnetic moment for a certain range of impurity 
concentrations \cite{Al2}; and (viii) (quantum) phase transitions 
driven by a change of the impurity concentration 
\cite{hfs,rechf,Al2}. 
Considering these similarities, it is tempting to speculate that our 
microscopic model supports the phenomenological Ginzburg-Landau 
approach of \cite{Mo} for the coexistence of superconductivity and 
magnetism in this class of heavy fermion alloys \cite{coh}. Still, 
the presence of possible nodes on the Fermi surface of unbound 
electrons in 2D or 3D may change the picture \cite{nodes}, as may 
the fact that in our model the mechanism for superconductivity lives 
outside the magnetic ions --- in contrast to the conventional 
picture of the U-based heavy fermion superconductors. It is 
interesting to note that while integrability imposes certain formal 
restrictions on possible hybridization terms, our particular choice 
in (2) shares some similarities with that recently proposed for 
U$_{1-x}$Th$_x$Be$_{13}$ on purely phenomenological grounds 
\cite{SAC}. This compound shows an admixture of {\em two} 
low-energy configurations of U ions: the magnetic state $\Gamma_6$ 
of the $5f^3$ configuration of U$^{3+}$ and the non-magnetic doublet 
$\Gamma_3$ of the $5f^2$ configuration of U$^{4+}$. This should be 
compared with the special case of $S={1\over2}$ in (2), with 
$\theta$ chosen to remove the degeneracy between magnetic and 
non-magnetic states: $\theta = 4[E(5f^3) - E(5f^2)]$, although in 
our case the magnetic state is four-fold degenerate, in contrast to 
\cite{SAC} where the magnetic state has a two-fold degeneracy.   

Results on the effect of {\em random} distributions of impurity 
parameters in spin-gapped electron systems, as well as a detailed 
analysis of the present model will be reported elsewhere.

The authors acknowledge financial support from the Swedish Institute 
(A.~A.~Z.) and the Swedish Natural Science Research Council (H.~J.).

\end{multicols}


\begin{references}
\bibitem{BT} A.~V.~Balatsky and S.~A.~Trugman, Phys. Rev. Lett. {\bf 
79}, 3767 (1997).
\bibitem{hf} H.~v.~L\"ohneysen {\em et al.}, Physica B {\bf 230--232} 
550(1997); B.~B.~Maple {\em et al.}, J. Low Temp. Phys. {\bf 99}, 223 
(1995);O.~O.~Bernal {\em et al.}, Phys. Rev. Lett. {\bf 75}, 2023 
(1995).
\bibitem{Lee} N.~Nagaosa and P.~A.~Lee, Phys. Rev. Lett. {\bf 79}, 
3755 (1997).
\bibitem{GM} For a recent review, see e.g. A.~M.~Goldman and 
N.~Markovi\'c, {\em Physics Today}, November, 1998, p. 39-44 and 
references therein.
\bibitem{sup} B.~T.~Mattias, H.~Suhl and E.~Corenzwit, Phys. Rev. 
Lett. {\bf 1}, 92 (1958); A.~A.~Abrikosov and L.~P.~Gor'kov, Sov. 
Phys. JETP {\bf 12}, 1243 (1961); P.~W.~Anderson, Phys. Rev. Lett. 
{\bf 3}, 325 (1959); D.~Markowitz and L.~P.~Kadanoff, Phys. Rev. {\bf 
131}, 563 (1963). 
\bibitem{hfs} B.~Bartlogg {\em et al.}, Phys. Rev. Lett. {\bf 55}, 
1319 (1985); A.~P.~Ramirez {\em et al.}, {\em ibid} {\bf 57}, 1072 
(1986); R.~H.~Heffner {\em et al.}, {\em ibid} {\bf 65}, 2816 (1990). 
\bibitem{rechf} F.~Kromer {\em et al.}, Phys. Rev. Lett., {\bf 81}, 
4476 (1998).
\bibitem{Al1} F.~G.~Aliev {\em et al.} Europhys. Lett., {\bf 32}, 765 
(1995). 
\bibitem{SS} For a review, see P.~Schlottmann and P.~D.~Sacramento,
Adv. Phys. {\bf 42}, 641 (1993), and references therein. 
\bibitem{Sch} P.~Schlottmann, Phys. Rev. Lett. {\bf 68}, 1916 (1992); 
Phys. Rev. B, {\bf 49}, 6132 (1994). 
\bibitem{Sch1} P.~Schlottmann, Z. Phys. B {\bf 59}, 391 (1985).
\bibitem{KBI} See, e.g. V.~E.~Korepin, N.~M.~Bogoliubov and 
A.~G.~Izergin {\em Quantum Inverse Scattering Method and Correlation 
Functions}, Cambridge University Press, 1993, and references therein. 
\bibitem{fin} P.~Schlottmann and A.~A.~Zvyagin, Phys. Rev. B {\bf 
56}, 13989 (1997); P.~Schlottmann, {\em ibid}, {\bf 57}, 10638 (1998).
\bibitem{TB} L.~A.~Takhtajan, Phys. Lett. {\bf A87}, 479 (1982); 
H.~M.~Babujian, Nucl. Phys. {\bf 215 B} [FS], 317 (1983).
\bibitem{hf1} See, e.g., C.~L.~Lin {\em et al.}, Phys. Rev. Lett. 
{\bf 58}, 1232 (1987); A.~Maeda {\em et al.}, {\em ibid} {\bf 74}, 
1202 (1995).
\bibitem{AM} F.~C.~Alcaraz and M.~J.~Martins, J. Phys. A: Math. Gen. 
{\bf 21}, 4397 (1988).
\bibitem{Fel} R.~Felten {\em et al.}, Europhys. Lett., {\bf 2}, 323 
(1986). 
\bibitem{SAC} A.~Schiller, F.~B.~Anders and D.~L.Cox, Phys. Rev. 
Lett, {\bf 81}, 3235 (1998). 
\bibitem{Ram} A.~P.~Ramirez {\em et al.}, Phys. Rev. Lett., {\bf 73}, 
3018 (1994). 
\bibitem{Al2} F.~G.~Aliev {\em et al.}, JETP Lett., {\bf 60}, 591 
(1994).
\bibitem{Mo} V.~V.~Moshchalkov, JETP Lett. {\bf 45}, 224 (1987).
\bibitem{coh} Note that the local character of the attraction for the 
electrons due to the exchange interaction in (\ref{host}) implies a 
short superconducting coherence length $\xi_0 \propto \pi v_F 
/\Delta$, which is consistent with the relatively short coherence 
lengths in unconventional superconductors, as compared with the 
ordinary BCS-like superconductors.
\bibitem{nodes} These nodes may e.g. stabilize an intermediate 
(spin-glass-like) phase between the antiferromagnetic and 
superconducting phases, as in the high-$T_c$ cuprates.
\end{references}
\end{document}